\documentstyle[aps,preprint]{revtex}

\begin{document}

\title{Derivation of the transverse force on a moving vortex in a superfluid }

\author {D.J. Thouless$^1$, P. Ao$^{2,1}$, and Q. Niu$^3$}

\address{{$^1$Department of Physics, Box 351560,
    University of Washington, Seattle, WA 98195, USA } \\
{$^2$Department of Theoretical Physics, 
                  Ume\aa \ University, Ume\aa, Sweden} \\
{$^3$Department of Physics, University of Texas, Austin, 
             TX 78712, USA } }

\maketitle

\abstract{We describe an exact derivation of the total 
nondissipative transverse force acting on a
quantized vortex moving in a uniform background. 
The derivation is valid for neutral boson or fermion superfluids,
provided the order parameter is a complex scalar
quantity. The force is determined by the one-particle
density matrix far away from the vortex core, and is found to be the 
Magnus force proportional to the superfluid 
density.\footnote{The work was supported by US NSF Grant No. DMR-9220733
and Swedish NFR(P.A.). }  }


Using the analogy to the classical fluid dynamics 
one finds that a moving vortex in a
superfluid feels a force proportional to its velocity ${\bf v}_V$ as
${\bf F}_M = \rho_s {\bf K} \times {\bf v}_V$.\cite{donnelly}
Here $\rho_s$ is the superfluid density and  ${\bf K}$ the quantized 
circulation.\footnote{to appear in the 
Czechoslovak Journal of Physics, Vol.46 (1996), Suppl. S1, PP 27}
In earlier work we have shown the close connection between the Magnus
force and a Berry phase \cite{ANT}, but our arguments
might depend  on a detailed model of the vortex structure.
This does not give a satisfactory answer to 
the question of  various possible corrections to this expression, 
as have been suggested in Ref.\cite{iordanskii,volovik}.
In the present paper we consider a
single, isolated vortex forced to move through an infinite superfluid with
uniform velocity ${\bf v}_V$ by some moving pinning potential.  We
are able to make an exact evaluation of the coefficient of ${\bf v}_V$ in
the force which has to be applied to the vortex in terms of the integral
of the momentum density round a large loop that surrounds the vortex.
The first step in the argument involves a
perturbative calculation of the effect of the motion of the vortex in
terms of the instantaneous eigenstates of the Hamiltonian.  In the case of
a system with no inhomogeneous substrate this expression can be written
in terms of a Berry phase \cite{geometric}.  This expression can be
written in terms of the derivatives of one- and two-particle Dirac density
matrices, or, alternatively, as the commutator of two components of the
total momentum operator.  This can be turned into an integral over a
surface (or a loop for the two-dimensional case) surrounding the vortex
and at a large distance from it.  From this we can conclude that there
are no corrections to $\rho_s {\bf K} \times {\bf v}_V$. 
   
For simplicity we work with a two-dimensional system, but the argument can
readily be applied to a three-dimensional system. We assume that the system
is homogeneous and infinite. The isolated vortex in the superfluid can be
pinned to a position ${\bf r}_0$ by applying a potential $\sum_iV({\bf
r}_0-{\bf r}_i)$ which acts on all the particles in the superfluid.
For a static problem this potential can be arbitrarily weak, but when
the pinning potential is made to move through the superfluid it must
have a strength sufficient that the vortex does not get detached from
the pinning center, either by tunneling with the aid of the Magnus
force, or by acquiring enough energy from the phonon system.  The
potential has to be strong enough to break the degeneracy of this
broken symmetry vortex state and allow us to use a perturbative
treatment of the velocity.  
The pinning potential may also be very strong, as it is in the Vinen 
experiment\cite{vinen}, where the circulation is pinned to a solid wire.  
Our results are quite independent of the details of the pinning
potential, provided that it is strong enough to allow perturbative
methods to be used, and does not break cylindrical symmetry.

In terms of the instantaneous eigenvalues $E_\alpha(t)$ and eigenstates
$|\Psi_\alpha(t)\rangle$, for which we choose phases such that
$\langle\Psi_\alpha|\dot\Psi_\alpha \rangle=0$, 
and letting $\nabla_0$ denotes the partial derivative with respect to
the position ${\bf r}_0$ of the pinning potential, we have 
the expectation value of the force 
on the pinning potential first order in vortex velocity as
\begin{equation}
   {\bf F}= \sum_\alpha f_\alpha 
            \left\langle \Psi_\alpha \left|
      \nabla_0H {i\hbar{\cal P}_\alpha\over E_\alpha- H} {\bf v}_V\cdot
        \nabla_0 +{\rm h.c.} \right| \Psi_\alpha \right\rangle \;.
\end{equation}
Since $\nabla_0H$ is the commutator of the operator $\nabla_0$
with $H$, the commutator cancels the energy denominator, and 
eq. (1) can be rewritten as
\[
   {\bf F} = - i\hbar\hat{\bf n} \times {\bf v}_V\sum_\alpha f_\alpha
     \left( \left\langle {\partial\Psi_\alpha\over \partial x_0} \left|
    {\partial\Psi_\alpha\over \partial y_0}\right\rangle \right. \right.
\]
\begin{equation}
    - \left. \left. \left\langle
    {\partial\Psi_\alpha\over \partial y_0} \right| {\partial\Psi_\alpha\over
    \partial x_0}\right\rangle  \right) \;,
\label{eq:berryphase}
\end{equation}
where $\hat{\bf n}$ is the unit vector normal to the plane.
The Berry phase associated
with a closed loop in ${\bf r}_0$ can be written as the integral
over the area enclosed by the loop of $ F/\hbar v_V$.
Equation (\ref{eq:berryphase}) corresponds to the familiar form
for the Berry phase \cite{geometric}. 
Since we can choose the wave functions in such a way that the
dependence on ${\bf r}_0$ is entirely through $({\bf r}-{\bf r}_0)$,
the partial derivatives with respect to ${\bf r}_0$ can be replaced
by a sum over partial derivatives with respect to the particle
coordinates ${\bf r}_j$.  Upon thermal average, this expression can now be 
written in terms
of the Dirac density matrices as
\[
  F /v_V =-i\hbar \hat{\bf n}\cdot \int d^2r[\nabla\times\nabla'\rho({\bf
     r}', {\bf r})]_{r=r'}
\]
\begin{equation}
  -i\hbar \hat{\bf n}\cdot\int d^2r_1\int d^2r_2 [2 \nabla_1
  \times \nabla'_2 \Gamma({\bf r}'_1,{\bf r}'_2; {\bf
  r}_1,{\bf r}_2)]_{r=r'} \;,  
\label{eq:magnusforce1}
\end{equation}
where $\rho$ and $\Gamma$ are the one- and two-particle Dirac
density matrices for the system, and the sum over $i$ and $j$ denotes
a sum over all the particles in the system. 

The integral over the two-particle density matrix $\Gamma$ vanishes.  
A formal way of getting the result is to say that the first
line of eq.\ (\ref{eq:magnusforce1}) gives the transverse force in terms
of the expectation value of the commutator of the $x$ and $y$
components of the total momentum of the particles.  This commutator is
a one-particle operator given by
$   [P_x,P_y]= \int\int dx\,dy\left({\partial \psi^{\dag}\over \partial x}
   {\partial\psi\over \partial y} -{\partial \psi^{\dag}\over \partial y}
   {\partial\psi\over \partial x} \right) ,
$
where the $\psi^{\dag},\psi$ are creation and annihilation operators
for fermions or bosons. Since it is a one-particle operator, its
expectation value is given by the one-particle Dirac density matrix.

The integrand of the first term in eq.\ (\ref{eq:magnusforce1}) 
can be written as half the curl of $(\nabla'
-\nabla)\rho({\bf r}',{\bf r})$ evaluated at ${\bf r}={\bf
r}'$. Now divide the integrals over ${\bf r}$ and ${\bf r}_1$ into a sum
over finite areas labeled with an index $\sigma$; the first of
these areas will be centered on the position of the vortex, and the
others are all well away from the vortex. Stokes' theorem can be used
to write the result as 
\begin{equation}
   F / v_V= \sum_\sigma\oint_\sigma {i\hbar\over 2}d{\bf r}
    \cdot[(\nabla- \nabla')\rho({\bf r}', {\bf r})]_{r=r'}  \;. 
\label{eq:magnusforce2}
\end{equation}
This result is exact, and there are no contributions to the
integrals from the neighborhood of the vortex core, since we have
chosen the boundaries of the regions of integration to be well away
from the core; any contributions from the core states or from the
properties of the core must be reflected in the density matrices
well away from the core. 
In particular, because localized states inside the 
vortex core, such as occur for an s-wave superconductor
\cite{caroli}, do not influence the one-particle density matrix far away
from the core, their contributions to the transverse force must be zero.
It should be pointed out that this property is not transparent in
eq.\ (\ref{eq:magnusforce1}), where both detailed forms of density matrices
and explicit integrations  are needed to recover it. 

For a neutral superfluid the integrand in eq.\ (\ref{eq:magnusforce2})
is just the momentum density, equal, by definition, to $\rho_s{\bf
u}_s+\rho_n{\bf u}_n$, where ${\bf u}_s$ and ${\bf u}_n$ are the local
values of the superfluid and normal velocities. If the circulation of
the normal fluid is zero, this term gives ${\bf F} = \rho_s {\bf
K}\times{\bf v}_V$, the Magnus force ${\bf F}_M$
\cite{donnelly,ANT,vinen}.


\begin{thebibliography}{99}
\bibitem{donnelly} 
     W.F. Vinen, in {\it Superconductivity}, II, Edited by R.D. Parks,
     Marcel Dekker, New York, 1969;
     R.J. Donnelly, {\it Quantized Vortices in Helium II},
           Cambridge University Press, Cambridge, 1991.
\bibitem{ANT}
   P. Ao and D.J. Thouless, Phys. Rev. Lett. {\bf 70}, 2158 (1993); 
   P. Ao, Q. Niu, and D.J. Thouless, 
     Physica {\bf  B194-196}, 1453 (1994); 
   Q. Niu, P. Ao, and D.J. Thouless, Phys. Rev. Lett., {\bf 72}, 1706 (1994).
\bibitem{iordanskii}
   S.V. Iordanskii, Sov. Phys. JETP {\bf 22}, 160 (1966);
  E.B. Sonin, Rev.\ Mod.\ Phys.\ {\bf 59}, 87 (1987).
\bibitem{volovik}
   N.B. Kopnin and M.M. Salomaa, Phys. Rev. {\bf B44}, 9667 (1991);
   G.E. Volovik, JETP Lett. {\bf 62}, 66 (1995).  
\bibitem{geometric}
   {\it Geometric Phases in Physics}, Edited by A. Shapere and F. Wilczek,
    World Scientific, Singapore, 1989. 
\bibitem{vinen} 
   W.F. Vinen,  Proc.\ Roy.\ Soc.\ (London) 
     {\bf A260}, 218 (1961).
\bibitem{caroli}
   C. Caroli, P.G. de Gennes, and J. Matricon, Phys. Lett. {\bf 9}, 307 (1964).
\end{thebibliography}
\end{document}